\documentstyle{article}

\font\tenrm=cmr10
\font\tenit=cmti10
\font\elevenbf=cmbx10 scaled\magstep 1
\font\elevenrm=cmr10 scaled\magstep 1
\font\elevenit=cmti10 scaled\magstep 1

\def\ltap{\ \raisebox{-.4ex}{\rlap{$\sim$}} \raisebox{.4ex}{$<$}\ }
\def\gtap{\ \raisebox{-.4ex}{\rlap{$\sim$}} \raisebox{.4ex}{$>$}\ }

\renewenvironment{thebibliography}[1]
 { \elevenrm
   \begin{list}{\arabic{enumi}.}
    {\usecounter{enumi} \setlength{\parsep}{0pt}
     \setlength{\itemsep}{3pt} \settowidth{\labelwidth}{#1.}
     \sloppy
    }}{\end{list}}

\begin{document}
\rightline{\vbox{\halign{&#\hfil\cr
&NTUTH-93-12\cr
&August 1993\cr}}}
%
\begin{center}
\vglue 0.6cm
{
 {\elevenbf        \vglue 10pt
               SEARCHING FOR \mbox{\boldmath $t\to ch^0$}
                OR \mbox{\boldmath $h^0\to t\overline c$} \\
               \vglue 3pt
               AT \mbox{\boldmath $e^+e^-$} LINEAR COLLIDERS
\\}
\vglue 1.0cm
{\tenrm GEORGE W. S. HOU \\}
\baselineskip=13pt
{\tenit Department of Physics, National Taiwan University \\}
\baselineskip=12pt
{\tenit Taipei, Taiwan 10764, R.O.C.\\}}


\end{center}


\vglue 0.6cm
{\elevenbf\noindent 1. Why?}
\vglue 0.4cm
\baselineskip=14pt
\elevenrm
It is quite legitimate for one to ask: Why? Are there $t$--$c$--Higgs
couplings in Nature? In the Standard Model (SM),
tree level flavor changing neutral couplings (FCNC)
of the Higgs boson are absent,
while at the one--electroweak--loop level, it is still strongly suppressed
by the G.I.M. mechanism.  In ``standard" two Higgs doublet models$^1$
(called Model I and II),
FCNC Higgs couplings are removed by design, and
again suppressed by the G.I.M. mechanism at the one--loop level,
since the charged Higgs contribution is rather
similar to the $W$ boson contribution diagramatically.

However, we now know that the top is rather heavy, and so is the Higgs
boson, which functions as the ``agent of mass". Their heaviness
suggests that they may hold the key to the question of mass, and
as new, undiscovered particles, their properties certainly call for thorough
examination.

It was recently pointed out$^{2,3}$ that, with more than one Higgs doublet,
it is possible to have {\elevenit tree level}
FCNC Higgs couplings that satisfy the stringent
low energy constraints, but are precisely the largest for the top quark.
Two generic types of models are given.
In Model III, which is basically a generalization of the
model of Cheng and Sher,$^4$
FCNC Higgs couplings are of the form
\begin{equation}
\Delta_{ij}^f \frac{\sqrt{m_i m_j}}{v} \bar f_i f_j h^0,
\end{equation}
where $v$ is the electroweak symmetry breaking scale,
$f = u$, $d$ and $e$, and $\Delta$ is of order one.
We have denoted the
neutral Higgs boson with FCNC coupling generically as $h^0$.
It can also be viewed as the lightest such Higgs of
phenomenological concern.
Note that Eq. (1) reduces to the usual $m_i/v$ for flavor diagonal
couplings, while the given pattern is consistent with the quark
mass and mixing hierarchies.$^2$

Model IV is constructed$^2$ as a hybrid of Models I/II and III
that eliminates FCNC Higgs couplings in the $d$ and $e$ type
fermion sectors, but
up type quarks have FCNC Higgs couplings of the form of Eq. (1).
It is clear that the $t$--$c$--$h^0$ coupling is the largest
possible one with 3 fermion generations in both Models III and IV.

Before one explores $t\to ch^0$ decays, it is prudent to check
with whatever low energy constraint that is available.
The FCNC Higgs couplings of Eq. (1) are proportional to the geometric mean of
the external fermion masses, and are thereby very suppressed
in low energy processes.$^4$
Nevertheless, the $d$ type quarks and charged leptons
provide stringent constraints, the most notable ones are
$K^0$--$\bar K^0$ mixing,  $B^0$--$\bar B^0$ mixing and
$\mu\to e\gamma$, and an earlier analysis$^5$ lead to a
not so stringent bound of $m_{h^0 } \gtap 80$ GeV. It was recently
pointed out$^6$ that a two--loop mechanism makes
$\mu\to e \gamma$ into a rather
effective constraint for Model III,
leading to a bound of $m_{h^0 } \gtap 200$ GeV.
Although still viable and interesting,
the heaviness of the Higgs bosons makes
phenomenology rather limited for the near future.
However, Model IV is distinguished in the sense that FCNC
couplings are decoupled from $d$ type quarks and charged leptons,
by construction. In fact, in Model IV, $t$--$c$--$h^0$ coupling
can be much larger than given by Eq. (1).
Given that the $D^0$--$\bar D^0$ mixing bound is rather poor,
there is practically no limit on $m_{h^0}$ coming from low energy
FCNC constraints! We remark that the present LEP limit on the neutral
Higgs boson mass, $m_{H^0} \gtap 60$ GeV, is weakened in general two
Higgs models.

The upshot is that, there is in fact little {\elevenit direct} experimental
constraint on $t$--$c$--$h^0$ couplings. If extra (more than one)
scalar bosons exist, FCNC couplings of order $\sqrt{m_c m_t}/v$
or greater are conservative and reasonable. Note that since
Model II is the Higgs sector of minimal SUSY (MSSM),
tree level FCNC is absent in MSSM.

We shall take $t$--$c$--$h^0$ coupling as
\begin{equation}
\lambda_{ct} = \sqrt{m_c m_t}/v = \sqrt{G_F m_c m_t}
\end{equation}
for illustration purposes.
For the mass range $m_t = 50$ -- $200$ GeV, one finds
\begin{equation}
\frac{\mbox{$\lambda_{ct}$}}{g_2} = \sqrt{m_c m_t}/M_W \simeq 0.1\ - \ 0.2,
\end{equation}
{\elevenit i.e.} weaker than the weak gauge coupling (of order $10$--$20\%$
in relative strength).
\vglue 0.5cm
{\elevenbf \noindent 2. Phenomenology}
\vglue 0.4cm
Phenomenological impact would depend on whether the top is lighter
than $h^0$ and ...
\vglue 0.2cm
{\elevenit\noindent 2.1.  $m_{h^0} < m_t < M_W + m_b$}
\vglue 0.1cm
One may think that CDF had ruled out this scenario long ago.
In fact, the 1992 (1989) limit $m_t \gtap 91$ GeV is now upgraded
by the new 92--93 run to roughly $m_t \gtap 110$ GeV by
CDF and D0.$^7$ Recall, however, that this is done by
assuming the SM $BR(t\to b\ell\nu) \simeq 1/9$ for $\ell = e,\ \mu$.
CDF and D0 in fact measure the effective cross section
in the $\ell\nu + jets$ and $\ell\ell^\prime\nu\nu + jets$
signatures. Define $B_{h^0} \equiv BR(t\to ch^0)$, {\elevenit i.e.}
\begin{equation}
B_{h^0} \equiv \frac{\Gamma (t\to ch^0)}
{\Gamma (t\to bW^{\left(*\right)}) + \Gamma (t\to ch^0)}
= \frac{R}{1+R},
\end{equation}
where $R$ is the ratio of the $t\to ch^0$ rate {\elevenit w.r.t.}
the SM $t\to bW^{\left(*\right)}$ rate.
The counting rate at the Tevatron in these modes should be
\begin{equation}
N(\ell\nu + jets) \cong
{\cal L}\times \sigma_{t\bar t}
\times \frac{8}{27} \times (1-B_{h^0})(1+\frac{1}{2} B_{h^0})
\times \epsilon,
\end{equation}
and
\begin{equation}
N(\ell\ell^\prime\nu\nu + jets) \cong
{\cal L}\times \sigma_{t\bar t}
\times \frac{4}{81} \times (1-B_{h^0})^2 \times \epsilon,
\end{equation}
where $\epsilon$ is the detector acceptance and efficiency
for the respective modes. It is clear that if $B_{h^0} \rightarrow 1$,
the $m_t$ limit would weaken.

One might expect that $B_{h^0}$ to be relatively small since from
Eq. (2) the $t$--$c$--$h^0$ coupling is weaker than the $SU(2)$ gauge
coupling. However, in this mass range, $t \to bW^*$ is a three body decay,
while the $t\to ch^0$ decay is a two body process,
and the smaller FCNC coupling in fact prevails.
With the $t$--$c$--$h^0$
coupling of Eq. (2), $R$ varies from more than 10 to order 1
as $m_t$ goes from $50$ GeV to $85$ GeV.$^2$
It can be
considerably larger in Model IV since the coupling is unconstrained.
Thus, $B_{h^0}$ could be from $50\%$ to more than $90\%$,
and the $m_t$ limit from CDF is considerably weakened.
Note that the limit from the usually
more stringent $\ell\ell^\prime\nu\nu + jets$ mode
weakens faster than the $\ell\nu + jets$ mode.

With more accumulated data in the 92--93 run, one could
weaken the assumptions on $B_{s.l.} \equiv BR(t\to \ell\nu + X)$
and explore the $B_{s.l.}$--$m_t$ plane. One would equivalently
be ``measuring" $B_{h^0}(m_t)$. This is apparently not yet done,
as the new searchs seem to be aimed directly at a heavier (than 91 GeV)
top, and tighter lepton $p_T$ cuts have been applied.$^7$
However, in case CDF and D0 do not
recheck the light top option, or if they simply
cannot rule out the possibility because $B_{h^0} \rightarrow 1$,
the top quark with $m_t \ltap M_W$ will show up readily
at LEP--II as it is turned on in a couple of years.
\vglue 0.2cm
{\elevenit\noindent 2.2.  $m_t > M_W + m_b$, $m_{h^0}$}
\vglue 0.1cm
As $m_t$ grows beyond the $M_W$ threshold, $B_{h^0}$ quickly drops
since $t\to bW$ becomes a two body process. The CDF/D0 search
assuming SM branching ratios would then be valid. However,
from Eq. (3), we see that in general $B_{h^0} \sim 1\%$,
{\elevenit i.e.} basically just the ratio of couplings involved in
the respective processes. This branching ratio stays rather flat until one
reachs the kinematic limit for $t\to ch^0$.$^2$

Although the top would be found by CDF/D0, searching for
$t\to ch^0$ at $1\%$ level would be impossible for the Tevatron,
and may be very difficult even for the SSC/LHC.
Here, we are interested in whether the 500 GeV $e^+e^-$ Linear Collider
could be of use. The typical benchmark is to assume 10 fb$^{-1}$ in
integrated luminosity, which corresponds to order of $10^4$ $t\bar t$
pairs.
The single most important aspect of Linear Collider capabilities
for this puprose is the $b$ tagging efficiency.
SLD has demonstrated $b$--tagging efficiencies at the $60$--$70\%$
level.$^8$ This is because of the very small beam spot size
and superb micro--vertex capabilities.
Let us consider two strategies.
Using $t\to \ell\nu + X$ as tag, one expects several thousand
{\elevenit clean} top events, where the event kinematics can be used to
project out 3 jets with $M_{jjj}$ consistent with $m_t$.
Assuming $b$ tagging efficiency to be order 0.5,
one expects a few events that are consistent with
$t\to ch^0\to cb\bar b$. The expected background from $t\to bW\to bc\bar b$
is 10 times smaller, but actual background rejection has to be further studied.
The second method is to demand {\elevenit three} $b$--jets. Again,
background suppression has to be studied, but one expects to
retain 10--30 $t\to cb\bar b$ events. However, the
six jet event signature is quite entangled, and much
work would be needed to be able to utilize the higher statistics.

Clearly, 100 fb$^{-1}$ would be more desirable.
Together with the available handles of $m_t$ known to 1 GeV,
and perhaps the knowledge of $m_{h^0}$.
the $b$--tagging capability at the Linear Collider certainly would
make the search for $t\to ch^0$ easier.

\vglue 0.2cm
{\elevenit\noindent 2.3.  $m_{h^0} > m_t$}
\vglue 0.1cm
The competing modes would be $h^0\to b\bar b$, $t\bar t$ and
$WW$, $ZZ$, but $h^0\to t\bar c$ may well be dominant,
especially if $m_t \ltap m_{h^0} \ltap 2m_t$, $2M_W$.
For example, for $m_{h^0} = 160$ GeV and $m_t = 130$ GeV,
the two allowed modes $h^0\to t\bar c$ and $b\bar b$ are comparable.
Since the top quark is presumably already found,
detection of $h^0\to t\bar c$ should be straightforward
at the Linear Collider.
Note, however, that Higgs production cross sections
(scalar, pseudoscalar, charged) are modified in multi--Higgs models.

\vglue 0.5cm
{\elevenbf \noindent 3. Summary \hfil}
\vglue 0.4cm
FCNC $t\to ch^0$ decays may not be as rare as Standard Model prescribes.
Such couplings are not ruled out by known physics, and since
the top and Higgs are new, heavy particles, we need to
thoroughly examine their properties. The $t\to ch^0$ mode
could dominate over $t\to bW^*$ and allow the top to
evade the CDF bound for the region $m_t \ltap M_W$.
The region can be covered readily by LEP--II.
For heavier top, the ballpark number is $BR(t\to ch^0)\sim 1\%$,
which cannot be studied at the Tevatron.
The expected high $b$--tagging efficiency can be used
at the 500 GeV Linear Collider to establish $t\to ch^0$ at
this level. However, 100 fb$^{-1}$ would be needed to have enough statistics
to study the decay in more detail.
In case $h^0$ is heavier than the top, searching for $h^0\to t\bar c$, which
could well be the dominant mode, should be easy.
The discovery of FCNC $t\to ch^0$ or $h^0\to t\bar c$ decays
would not only be very exciting in itself, it would
also rule out the minimal supersymmetric standard model.
\vglue 0.5cm
{\elevenbf \noindent 4. Acknowledgements \hfil}
\vglue 0.4cm
I thank M. Peskin for drawing to my attention the $B$ tagging
efficiency at Linear Colliders.
\vglue 0.5cm
{\elevenbf\noindent 5. References \hfil}
\vglue 0.4cm

\end{document}